\documentclass[prd,aps,twocolumn,amsmath,amssymb,nofootinbib,preprintnumbers]
{revtex4}

\usepackage{graphicx}
\usepackage{dcolumn}
\usepackage{bm}
\usepackage{amsmath}
\usepackage{amsfonts}
\usepackage{euscript,bbm}
\usepackage{ifthen}
\usepackage{psfrag}
\usepackage{slashed}

\def\ls{\mathrel{\lower4pt\vbox{\lineskip=0pt\baselineskip=0pt
           \hbox{$<$}\hbox{$\sim$}}}}
\def\gs{\mathrel{\lower4pt\vbox{\lineskip=0pt\baselineskip=0pt
           \hbox{$>$}\hbox{$\sim$}}}}
\def\drawbox#1#2{\hrule height#2pt

\hbox{\vrule width#2pt height#1pt \kern#1pt
              \vrule width#2pt}
              \hrule height#2pt}

\def\Asym#1#2{\vcenter{\vbox{\drawbox{#1}{#2}
              \kern-#2pt       
              \drawbox{#1}{#2}}}}

\newcommand{\Expect}[1]{\left\langle #1 \right\rangle}
\newcommand{\be}{\begin{equation}}
\newcommand{\ee}{\end{equation}}
\newcommand{\bea}{\begin{eqnarray}}
\newcommand{\eea}{\end{eqnarray}}

\begin{document}

%
\title{Affleck-Dine Baryogenesis in Effective Supergravity}

\author{Bhaskar Dutta$^{1}$}
\author{Kuver Sinha$^{1}$}

\affiliation{$^{1}$~Department of Physics, Texas A\&M University, College Station, TX 77843-4242, USA}


\begin{abstract}

We investigate the viability of Affleck-Dine baryogenesis in $D=4, N=1$ supergravity descending from string theory. The process relies on an initial condition where visible sector supersymmetric flat directions obtain tachyonic masses during inflation. We discuss this condition for a variety of cases where supersymmetry is broken during inflation by a geometric modulus or hidden sector scalar, and outline scenarios where the initial condition is satisfied.

\end{abstract}
MIFPA-10-33
\\ July, 2010
\maketitle


\section{Introduction}

$M$ theory and its weakly coupled string theory limits contain numerous moduli, whose vacuum expectation values determine the masses and coupling constants of the low energy theory. Along with moduli stabilization \cite{Douglas:2006es}, there has been much focus on the effects of these scalar fields on cosmology. Certainly, inflation offers the most promising field of application and has also been the most studied \cite{McAllister:2007bg}, \cite{Baumann:2009ni}, but one also has other paradigms such as baryogenesis.

Most standard processes such as electroweak baryogenesis or leptogenesis are relatively independent of a particular UV completion. However, the situation is different for Affleck-Dine baryogenesis \cite{Dine:1995kz}, \cite{Dine:1995uk}, \cite{Affleck:1984fy},  which relies on an inflationary sector to produce coherent oscillations along a supersymmetric flat direction. The interaction between the inflationary sector and the flat direction occurs through Planck-suppressed operators whose form is critical for the success of the process. The setting is effective $D=4, N=1$ supergravity. It is natural to probe the success of the method when the low energy supergravity descends from string theory.

In this paper, we investigate initial conditions for Affleck-Dine (AD) baryogenesis in a variety of supergravity scenarios. Our purpose is two-fold. Firstly, as mentioned before, this offers an investigation of stringy effects in a standard method of baryogenesis.

Secondly, AD baryogenesis is very useful in fortifying the baryon asymmetry of the universe against late-time entropy production by moduli. The late decay of gravitationally coupled moduli leads to various modifications of standard cosmological scenarios such as the non-thermal production of dark matter \cite{Dutta:2009uf}, \cite{Acharya:2009zt} and baryogenesis \cite{Allahverdi:2010im}. In the early universe the moduli are displaced from the minimum of their potential and start oscillating. These coherent oscillations behave like non-relativistic matter once the Hubble expansion rate drops below their mass. Since the moduli couple to other fields only gravitationally, they are long-lived and can dominate the energy density of the universe. Moduli with masses above $20$ TeV decay before Big-Bang Nucleosynthesis (BBN); however, if they are sufficiently light (typically below $10^4$ TeV) they result in a very low reheat temperature, which is below a GeV. The decay of the modulus generates a large amount of entropy, which dilutes any baryon asymmetry that was created in a previous era (the dilution factor may be as large as $\sim 10^9$~\cite{Kawasaki:2007yy}). Generating sufficient baryon asymmetry below a GeV is a challenging task since sphaleron transitions are exponentially suppressed. The Affleck-Dine mechanism can produce an $ \mathcal{O}(1)$ baryon asymmetry and yield the desired value after late-time dilution by modulus decay.

A crucial element in concrete realizations of Affleck-Dine baryogenesis is to obtain a negative Hubble-induced mass term along the potential of visible sector supersymmetric flat directions. Thus, realizing the initial conditions for baryogenesis amounts to getting tachyonic soft masses along certain chiral fields due to supersymmetry breaking induced by the vacuum energy. If the vacuum energy during inflation is dominated by a field $\sigma$, this places constraints on the Kahler coupling between $\sigma$ and the visible sector. We study two broad choices for $\sigma$: when it is a geometric modulus of the compactification, and when it is a hidden sector scalar field. We find that in the case of geometric moduli in a type IIB compactification, tachyonic masses may be obtained if $\sigma$ is a local modulus of the visible sector. For a hidden sector scalar $\sigma$, gravitational couplings to the visible sector can induce tachyonic masses. To avoid tachyons in the final soft terms, $\sigma$ should have negligible F-term in the stable vacuum of the theory. We outline inflationary scenarios where $\sigma$ dominates the energy density during inflation but not in the final vacuum.   We also give the conditions for obtaining baryogenesis in the case of the weakly coupled heterotic string.

The rest of the paper is arranged as follows. In Section II, we outline the AD baryogenesis process. In Section III, we write down expressions for Hubble-induced mass terms for chiral superfields in $D=4,\, N=1$ supergravity. In Section IV, we study AD baryogenesis for the case when the energy density during inflation is dominated by a geometric modulus. In Section V, we study Hubble-induced masses due to a hidden sector matter field. We end with our conclusions.


\section{Affleck-Dine Baryogenesis}

We first outline the essential physics of the Affleck-Dine process. The mechanism depends on a very generic property of supersymmetric field theories: the existence of flat directions, which are unlifted in the absence of supersymmetry breaking, at the level of renormalizable operators. A flat direction is the normalized scalar component of a composite gauge invariant operator formed from the product of chiral superfields. For example, for $H_uL$, the flat direction $\phi$ is given by
\be
H_u \, = \, \frac{1}{\sqrt{2}}\left(
\begin{matrix}
0 \\
\phi \\
\end{matrix}
\right )\, ,
\,\,\,\,
L \, = \, \frac{1}{\sqrt{2}}\left(
\begin{matrix}
\phi \\
0 \\
\end{matrix}
\right )
\ee
In the very early universe, if conditions are met such that the flat direction $\phi$ is initially displaced from its true minimum, it starts to oscillate when the Hubble constant becomes smaller than the effective mass $V(\phi)^{\prime \prime} \sim m_{3/2}$. The energy of the oscillations corresponds to a condensate of non-relativistic particles. It is possible to store baryon number in a condensate in the particular case where the supersymmetric theory is the MSSM. After oscillations set in, a net baryon asymmetry may be produced depending on the magnitude of baryon number-violating terms in $V(\phi)$.

The finite energy density of the universe during inflation breaks supersymmetry and induces a SUSY breaking mass along $\phi$. If this Hubble-induced mass is tachyonic, the field is able to acquire a large vev during inflation. It then remains critically damped, and tracks an instantaneous minimum as long as $V(\phi)^{\prime \prime} \sim H^2 \gg m_{3/2}^2$. After $H \sim m_{3/2}$, it begins to oscillate.

MSSM flat directions are lifted by non-renormalizable terms in the superpotential
\be
W \, = \, \frac{\lambda}{nM_P^{n-3}}\phi ^n \,\, .
\ee
The potential along $\phi$, taking into account supersymmetry breaking terms due to the finite energy during inflation, is
\bea
V(\phi) &=& (c_H H^2 + m_{\rm soft}^2)|\phi|^2 + \left( \frac{(A+a_H H)\lambda \phi ^n}{n M_P^{n-3}} + {\rm h.c.} \right) \nonumber \\
&+& |\lambda|^2 \frac{|\phi | ^{2n-2}}{M_P^{2n-6}} \,\, .
\eea
Here, $c_HH^2$ and $a_HH$ denote soft parameters induced by inflation, while $m_{\rm soft}$ and $A$ arise from a supersymmetry breaking sector at the end of inflation. Note that $c_H$ could be of either sign.

For $H \gg m_{\rm soft}$, the curvature along $\phi$ is dominated by the Hubble-induced mass. If $c_H > 0$, the field sits at the origin and a condensate does not develop. However, if $c_H < 0$, the minimum lies at
\be
|\phi | \sim \left(\frac{\sqrt{-c}H M_P^{n-3}}{(n-1)\lambda}\right)^{\frac{1}{n-2}}
\ee
and the field tracks this minimum until $H \sim m_{\rm soft}$. The inclusion of the Hubble-induced A-term $a_HH$ gives $n$ discrete vacua in the phase of $\phi$, and the field settles into one of them. When $H \sim m_{\rm soft}$ the field begins to oscillate around the new minimum $\phi = 0$; thereafter the soft $A$-term becomes important and the field obtains a motion in the angular direction to settle into a new phase. The baryon number violation thus becomes maximal during this time and imparts asymmetry to the condensate. The final baryon to entropy ratio depends on the resulting baryon number per condensate particle, the total energy density in the condensate, and the inflaton reheat temperature. The baryon asymmetry is obtained as
\be \label{eta}
\frac{n_B}{n_{\gamma}} \, \sim \, 10^{-10} \left(\frac{T_{r, {\rm inflaton}}}{10^9\,{\rm GeV}}\right) \left(\frac{M_P}{m_{3/2}} \right)^{\frac{n-1}{n+1}}  \,\,.
\ee
Depending on $n$, the baryon asymmetry can be large which is very useful for models with lighter moduli mass $\sim 10^3$ TeV corresponding to low reheat temperature. There will be an additional dilution factor $\sim 10^{-8}$ which would allow us to obtain the correct amount of baryogenesis as observed in these models.

The initial condition in the above scenario is that $\phi$ is displaced from the origin to begin with. The sign of the Hubble-induced mass term $c_HH^2$ will depend on the coupling between the field $\sigma$ that dominates the energy during inflation and the flat direction $\phi$. Since the important couplings occur from Planck scale operators, supergravity interactions should be included. We now turn to the question of how to realize this initial condition.

\section{Hubble-Induced Mass Terms in Supergravity}

In this section we study the coupling between the MSSM flat direction and the Hubble mass-inducing field $\sigma$ in four dimensional effective supergravity.

The scalar potential is given by
\be
V \, = \, e^K\left(K^{i\overline{j}}D_{i}W D_{\overline{j}}\overline{W} - 3 |W|^2 \right)
\ee
The Kahler potential and superpotential can be written as
\bea \label{KW}
K &=& \widehat{K}(T_i,\overline{T}_i) + \widetilde{K}_{\alpha \overline{\beta}}(T_i,\overline{T}_i) \overline{\phi}^{\alpha}  \phi^{\beta}+ \ldots \nonumber \\
W &=& \widehat{W}(T_i) + \frac{1}{6}Y_{\alpha \beta \gamma} \phi^{\alpha \beta \gamma} \,\,.
\eea
In the above, $\phi$ denotes an MSSM chiral superfield, and $T$ is a generic modulus or hidden sector field. $\alpha, \, \beta, \, \gamma$ are flavor indices, and for the sake of simplicity we will henceforth consider diagonal matrices in flavor space. We will also use $\phi$ to denote the scalar component of the visible sector superfield, as well as a flat direction formed from combinations of chiral superfields.

Plugging $K$ and $W$ into the scalar potential and expanding in a series in $\overline{\phi}\phi$, one obtains the soft mass term for the chiral fields \cite{Brignole:1993dj}, \cite{Kaplunovsky:1993rd}
\be \label{msoft}
m_{\rm soft}^2 \, = \, m_{3/2}^2 + V_0 - F^{i}F^{\overline{j}}\partial_{i}\partial_{\overline{j}}\ln \widetilde{K} \,\,.
\ee
In the above, $V_0$ is the potential along the modulus, given by
\be
V_0 \, = \, F^{i}F^{\overline{j}}\widehat{K}_{i\overline{j}} - 3 m_{3/2}^2 + V_D
\ee
where $F^{i} = e^{\widehat{K}/2}D_{\overline{j}}\widehat{K}^{i\overline{j}}$ and $m_{3/2}^2 = e^{\widehat{K}} |W|^2$. We have also included possible D-term contributions to the vacuum energy. Note that in the effective low-energy theory the chiral matter is normalized as $\phi_{\rm normalized} \, = \,  \widetilde{K}^{1/2} \phi$. We will implicitly assume normalized fields henceforth.

Usually the soft mass computation due to modulus mediation proceeds by stabilizing moduli in the potential $V_0$ and using the values of the F-terms at the minimum with $V_0 \sim 0$. However, we want the soft masses induced due to the positive energy during inflation. Thus, we take the limit when $V_0 \sim F^{\sigma}F^{\overline{\sigma}}\widehat{K}_{\sigma\overline{\sigma}} + V_D$ is large and positive, and the energy is dominated by a combination of D-term effects and the F-term of a field $\sigma \in \{T_i\}$ during inflation.

We thus obtain
\be
c_H \, = \, \frac{m^2}{H^2} \, \sim \, 1 - \widehat{K}^{\sigma\overline{\sigma}}\partial_{\sigma}\partial_{\overline{\sigma}} \ln \widetilde{K} + \frac{V_D}{V_0}\widehat{K}^{\sigma\overline{\sigma}}\partial_{\sigma}\partial_{\overline{\sigma}} \ln \widetilde{K}\,\,.
\ee
It is clear that successful AD baryogenesis depends on several factors: $(i)$ geometric data: the Kahler potential $\widehat{K}$ of the moduli and the Kahler metric $\widetilde{K}$ of the chiral superfields in the visible sector $(ii)$ which particular field $\sigma$ dominates the energy density of the universe during inflation and $(iii)$ the importance of D-term effects relative to the vacuum energy during inflation .

Making definitive statements about the Kahler potential in effective supergravity is difficult since it is not protected by the non-renormalization theorems. One can assume various forms for the Kahler metric $\widetilde{K}$ of visible sector chiral matter which gives the coupling of flat directions to a modulus inflaton. For example, for minimal supergravity, $\widetilde{K} = const.\,$, and the Hubble-induced mass is positive, as is well known \cite{Casas:1997uk}, \cite{Kasuya:2006wf}.

We now study Hubble-induced mass terms for various cases. The remainder of the paper is divided into cases where the energy density is dominated by a geometric modulus, and cases where the energy density is dominated by hidden sector matter fields.

\section{Geometric Modulus Domination}

In this section, we consider Hubble-induced mass terms for cases when a geometric modulus dominates the energy density of the universe during inflation. For concreteness, we first consider candidates in type IIB string theory.

\subsection{Type IIB}

In a typical compactification in type IIB, the effective four-dimensional $N=2$ action (which is subsequently broken to $N=1$ by orientifold projections) consists of $h^{1,1} + 1$ hypermultiplets. One of them is the axio-dilaton. The bosonic components of the remaining $h^{1,1}$ consist of $(i)\,\,$ the volume modulus and other size moduli, denoted by $T_i$ and $(ii)\,\,$ the axionic modluli $b_i, c_i$ corresponding to integrating Neveu-Schwarz and Ramond-Ramond gauge potentials over non-trivial two-cycles. These moduli fields are natural condidates for the hidden sector that is responsible for inflation.

In a moduli stabilization scheme like KKLT, the complex structure moduli and the dilaton are fixed by fluxes. The superpotential is given by $\widehat{W} = W_{\rm flux} \, + W_{np}$, where $W_{\rm flux} = \int G_3 \wedge \Omega$ and the non-perturbative superpotential $W_{np}$ is sourced by gaugino condensation on D7-branes. The $T_i$ are fixed at an AdS vacuum by the non-perturbative contribution to the superpotential. The vacuum is then further lifted to a near Minkowski vacuum by D-term effects or by the introduction of anti-D3 branes. We denote these effects by $V_D$.

In many models of inflation that are based on KKLT moduli stabilization, the scale of inflation is typically set by the uplifting effect to avoid decompactification of the internal dimensions \cite{Kallosh:2004yh, Badziak:2008yg, Allahverdi:2009rm}. One thus obtains
\bea
V_D &\sim & m_{3/2}^2 \sim V_0 \,\, \nonumber \\
&\Rightarrow & c_H \, \sim 2 - 3 \widehat{K}^{\sigma\overline{\sigma}}\partial_{\sigma}\partial_{\overline{\sigma}} \ln \widetilde{K} \,\, .
\eea
It has been argued recently, however, that it may generically be possible to obtain models with $V_0 \gg m_{3/2}^2 \sim V_D$ \cite{He:2010uk}, by considering situations where $\Expect {W}$ relaxes from large to small values during inflation. For a generic scenario where inflation is dominated by F-terms of a field and other energy scales are much smaller, one obtains
\be \label{master}
c_H \, \sim 1 -  \widehat{K}^{\sigma\overline{\sigma}}\partial_{\sigma}\partial_{\overline{\sigma}} \ln \widetilde{K} \,\, .
\ee
For the rest of the paper, we will be interested in this general scenario where inflation is dominated by F-terms.

The Kahler potential for the moduli may be written, up to $\alpha^{\prime}$ corrections, as
\be \label{Kvol}
\widehat{K} \, = \, -2 \ln \, \mathcal{V} 
\ee
In the above, $\mathcal{V}$ denotes the Einstein-frame volume of the Calabi-Yau manifold.

The moduli dependence of the Kahler coupling $\widetilde{K}$ to chiral matter can be argued out on physical grounds \cite{Aparicio:2008wh}, \cite{Conlon:2006tj}, \cite{Conlon:2006wz}. For concreteness, we consider Calabi-Yau manifolds with $h^{1,1} \geq 2$ with all Kahler moduli stabilized \cite{Denef:2004dm}. The visible sector is taken to be localized near a small four-cycle $\tau_s \, = \, {\rm Re}\, T_s$. Due to holomorphy and the shift symmetry of ${\rm Im} \, T$, the Kahler moduli $T_i$ cannot appear at any level in perturbation theory in $W$, and hence, importantly, in $Y$ in Eq.~(\ref{KW}). The normalized Yukawa coupling
\be \label{YK}
\widehat{Y}_{\alpha \beta \gamma}(\tau_s, \mathcal{U}) \, = \, e^{\widehat{K}/2}\frac{Y_{\alpha \beta \gamma}(\mathcal{U})}{\left( \widetilde{K}_{\alpha} \widetilde{K}_{\beta} \widetilde{K}_{\gamma}\right)^{\frac{1}{2}}}
\ee
should only depend on local geometric data $\tau_s$ and complex structure moduli $\mathcal{U}$, but not the overall volume.

Thus, from Eq.~(\ref{YK}), we obtain
\be \label{Ktilde}
\ln \widetilde{K} \, = \, \frac{1}{3} \widehat{K} \, + \, \ln k(\tau_s, \mathcal{U}) \,\,,
\ee
where $k(\tau_s, \mathcal{U})$ is an undetermined, model-dependent function of local data.

Several cases are possible from Eq.~(\ref{master}) and Eq.~(\ref{Ktilde}).

$(i)$ The energy density is dominated by a modulus that is not a local modulus of the visible sector. In this case, one obtains
\be
c_H \, = \, \frac{2}{3} \,\,.
\ee
As an example of this case, one can consider a typical compactification on $\mathbb{P}^4_{[1,1,1,6,9]}$. There is a large four-cycle $\tau_b$ and a small four-cycle $\tau_s$ on which the visible sector is located. The volume is given by $\mathcal{V} = \tau_b^{3/2} - \tau_s^{3/2}$. If the large modulus $\tau_b$ dominates inflation, then one obtains positive $c_H$.\footnote{ In models with a single Kahler modulus, the calculation is slightly different, although the conclusion is similar. One has $\mathcal{V} = \tau_b^{3/2}$. If the visible sector is localized on a stack of D$7$-branes wrapped on a single $4$-cycle, then the physical Yukawas may be argued to scale as $\tau_b^{-1/2}$ and finally one obtains
\be \label{Ktilde1para}
\widetilde{K} \, = \, \frac{\tau_b^{\frac{1}{3}}}{\mathcal{V}^{2/3}} k_1(\mathcal{U}) \, \Rightarrow \, c_H \, = \, \frac{7}{9} \,\,.
\ee
}

A similar value of $c_H$ is obtained if Neveu-Schwarz or Ramond-Ramond axions dominate the energy density of the universe during inflation. For example, in compactifications with $O3/O7$ planes with $h_+^{1,1} = 1$, one obtains
\be \label{Vaxion}
\widehat{K} \, = \, -3 \ln \left( T_i + \overline{T_i} + \kappa^{ijk} b_jb_k  \right)
\ee
where $j,k = 1 \ldots h_{-}^{1,1}$ and $\kappa^{ijk}$ are triple intersection numbers \cite{McAllister:2008hb}. Following Eq.~(\ref{master}), Eq.~(\ref{Kvol}), and Eq.~(\ref{Ktilde}), it is clear that if the energy density is dominated by axions $b$, one obtains $c_H = 2/3$.\footnote{A similar conclusion holds in models of warped brane inflation \cite{Kachru:2003sx}, if the energy density during inflation is dominated by the brane-anti-brane separation.}

$(ii)$ The second case to consider is when the energy density is dominated by the F-term of a local modulus. In this case,
\be
c_H \, = \, \frac{2}{3} -  \widehat{K}^{T_s \overline{T_s}}\partial_{T_s}\partial_{\overline{T_s}} \ln \, k(\tau_s, \mathcal{U}) \,\, .
\ee
Obtaining an induced tachyonic mass imposes conditions on $\widehat{K}^{T_s \overline{T_s}}$ which depends on the specific Calabi-Yau compactification, and on $k(\tau_s, \mathcal{U})$, which depends on the construction of the visible sector. Thus, information both about the global compactification, as well as the local model, are required.

There are some general statements that can be made, however. The condition may be recast as
\be \label{cond1}
\partial_{T_s}\partial_{\overline{T_s}} \ln \, k(\tau_s, \mathcal{U}) \, > \, \frac{2}{3} \widehat{K}_{T_s \overline{T_s}} \,\, .
\ee
Since $\widehat{K}^{T_s \overline{T_s}} \, > \, 0$, it is necessary (but not sufficient) that
\be \label{localcondition}
\partial_{T_s}\partial_{\overline{T_s}} \ln \, k(\tau_s, \mathcal{U}) \, > \, 0 \,\,.
\ee
This condition should hold regardless of the global details of the compactification.

The exact nature of the function $k$ is difficult to determine. If the visible sector construction is in the supergravity limit, for example on intersecting stacks of magnetized $D7$ branes wrapping cycles $\tau_s$ larger than string scale in a localized region of the Calabi-Yau, then we may take $k(\tau_s, \mathcal{U}) \, \sim \, \tau_s^pk(\mathcal{U})$. This dependence holds for dilute flux where $\tau_s^{-1}$ controls the gauge coupling in the weak limit. Depending on the details of the construction of the matter fields, one has $0 \, < p \, < 1$ and in this class of constructions Eq.~(\ref{localcondition}) is violated. \footnote{It may be possible to arrange fluxes such that the next-to-leading-order term $k(\tau_s, \mathcal{U}) \, \sim \, \tau_s^{p-1} k_2(\mathcal{U})$ dominates over the leading term $\tau_s^{p}$. In that case, one obtains $c_H \sim 2(p - 1)/3 < 0$.}

On the other hand, if the visible sector gauge theory is constructed with branes at singularities, it is more difficult to write down supersymmetry breaking terms in general and the Hubble-induced mass in particular. While the dictionary between local geometry and superpotential deformations is quite well understood, the dictionary for SUSY breaking deformations requires understanding Kahler deformations and is much less clear.

Quiver gauge theories typically occur in non-geometric phases of the Kahler moduli space and the supergravity approximation becomes invalid \cite{Diaconescu:2007ah}. Using homological mirror symmetry, it is possible to map the type IIB brane configuration on a Calabi-Yau $X$ to D6-branes wrapping special Lagrangian cycles in type IIA on a mirror Calabi-Yau $Y$. The dynamics can be controlled if we work near the large complex limit point in the complex structure moduli space of $X$, which corresponds to the geomtric limit of the Kahler moduli space of $Y$ \cite{Diaconescu:2006nk}.

The problem of obtaining Hubble-induced supersymmetry breaking terms should thus be formulated in the language of type IIA string theory, where obtaining models of inflation is difficult \cite{Flauger:2008ad}. We will simply give an outline of how it may be possible to obtain tachyonic Hubble-induced masses. The holomorphic coordinates on the complex structure moduli space $\mathcal{M}$ of $Y$ may be identified with the dilaton and Kahler coordinates $\tau$ of $X$ by mirror symmetry. A priori, these are defined near the large complex structure limit of $\mathcal{M}$, but may be defined near the Landau-Ginzburg point (where our quiver theory is located) by analytic continuation. The moduli dependence of the Kahler metric $\widetilde{K}_{\mathcal{M}}$ for matter fields may be argued from the locality condition of the Yukawas, and the function $k$ expanded in a power series in $\tau$, obtaining
\be
k(\tau) \, = \, k_0 + k_1 \tau^p + \ldots \,\, .
\ee
where $k_0$ and $k_1$ are functions of the Kahler moduli of $Y$. Eq.~(\ref{cond1}) then imposes a relation on $k_0$ and $k_1$, which may be obtained by tuning even Ramond-Ramond fluxes on $Y$. In such a scenario, it is possible to obtain a negative value of $c_H$.

\subsection{Other Scenarios}

We now comment on some other scenarios of modulus domination.

In compactifications of $M$ theory on $G_2$ manifolds, the complexified moduli space has holomorphic coordinates $z_i = \theta_i + \imath \tau_i$, where $\theta_i$ are the axionic partners of the moduli $\tau_i$. In the fluxless sector, the moduli are stabilized by non-perturbative effects, sourced by strong gauge dynamics \cite{Acharya:2001gy}.

A family of Kahler potentials that are consistent with $G_2$-holonomy and widely used \cite{Acharya:2006ia} is
\bea
\widehat{K} \, &=& \, -3 log (4\pi^{1/3}\mathcal{V}) \nonumber \\
\mathcal{V} &=& \prod \tau_i^{a_i}, \,\, , \,\, \sum a_i = 7/3 \,\,.
\eea
The Kahler metric for chiral matter may be obtained similarly to the type IIB case \cite{Acharya:2008hi}, and if the moduli $\tau_i$ dominate the energy density during inflation, we get $c_H = 2/3$.

The weakly coupled heterotic string furnishes a canonical example of global visible sector model building. It is interesting to probe initial conditions for AD baryogenesis for effective supergravity theories derived from orbifold compactifications of the weakly coupled heterotic string \cite{Binetruy:1996nx}, \cite{Gaillard:2007jr}, \cite{Kane:2003iq}. In the linear superfield formalism, the Kahler potential is given by
\be \label{Khet1}
K \,=\, \ln(l) + g(l) -\sum\nolimits_i \ln x_i + \sum\nolimits_A X_A \,\, ,
\ee
where
\be \label{Khet2}
	x_i=T_i+\overline{T}_i-\sum\nolimits_A|\psi_{Ai}|^2, \qquad X_A=\left(\prod\nolimits_i x_i^{n_i^A}\right) |\psi_A|^2\,\,.
\ee
The $\psi_{Ai}$ are untwisted matter fields and $\psi_A$ are twisted matter fields with modular weight $n_I$. The $T_i$ are the three Kahler moduli of the compactification and $l$ is the dilaton. The function $g(l)$ is a non-perturbative contribution that stabilizes the dilaton.

The perturbative superpotential is given by
\be \label{Whet}
	W_p=\sum\nolimits_m \lambda_m \left[\prod\nolimits_i \eta(t_i)^{-2}\right] \prod\nolimits_\alpha \psi_\alpha ^{p_m^\alpha} \prod\nolimits_j \eta(t_j)^{2 p_m^\alpha q_j^\alpha},
\ee
where $\alpha$ denotes twisted as well as untwisted sector matter, running over $Ai,A$. The $\lambda_m$ are constants, the $p_m^\alpha$ are nonnegative integers and $\eta(t_I)$ is the Dedekind eta function.

%
In addition, there may be non-perturbative contributions.

If the energy density during inflation is dominated by the F-term of the overall Kahler moduli $T$, then from Eq.~(\ref{Khet1}), Eq.~(\ref{Khet2}) and Eq.~(\ref{master}), the Hubble-induced mass is
\be
c_H = \frac{m^2}{H^2} = 1 + \frac{n_{\alpha}}{3} \,\, ,
\ee
where $n = \sum n_i$. One thus requires the modular weight of the flat direction chiral field to be $n \, < \, -3$ for successful AD baryogenesis.

For Abelian orbifolds, the range of overall modular weights is given by $-3 \, \leq n_{Q,\overline{u},\overline{e}} \, \leq 0$ and $-5 \, \leq n_{L,\overline{d},H} \, \leq 1$ and more negative values are possible for higher choice of Kac-Moody levels \cite{Ibanez:1992hc}. Thus, for example, a flat direction formed by the gauge invariant operator $LL\overline{d}\overline{d}\overline{d}$ can have the correct modular weight for AD baryogenesis.


\section{Hubble-Induced Masses From Hidden Sector Matter Fields}

If supersymmetry breaking during inflation is dominated by the F-term of a hidden sector matter field, it is possible to obtain tachyonic Hubble-induced masses for visible sector flat directions. Broadly, some conditions should be satisfied for such a scenario (Figure 1).

$(i)$ Planck suppressed operators mixing the visible and inflationary sectors in the Kahler potential induce negative masses by gravity mediation along flat directions if the dimensionless coupling is chosen appropriately.

$(ii)$ The contribution to soft masses from the hidden matter sector in the final stabilized vacuum at the end of inflation should be negligible.

$(iii)$ The inflationary dynamics should be compatible with moduli stabilization.

\begin{figure}[ht]
\centering
\includegraphics[width=3.5in]{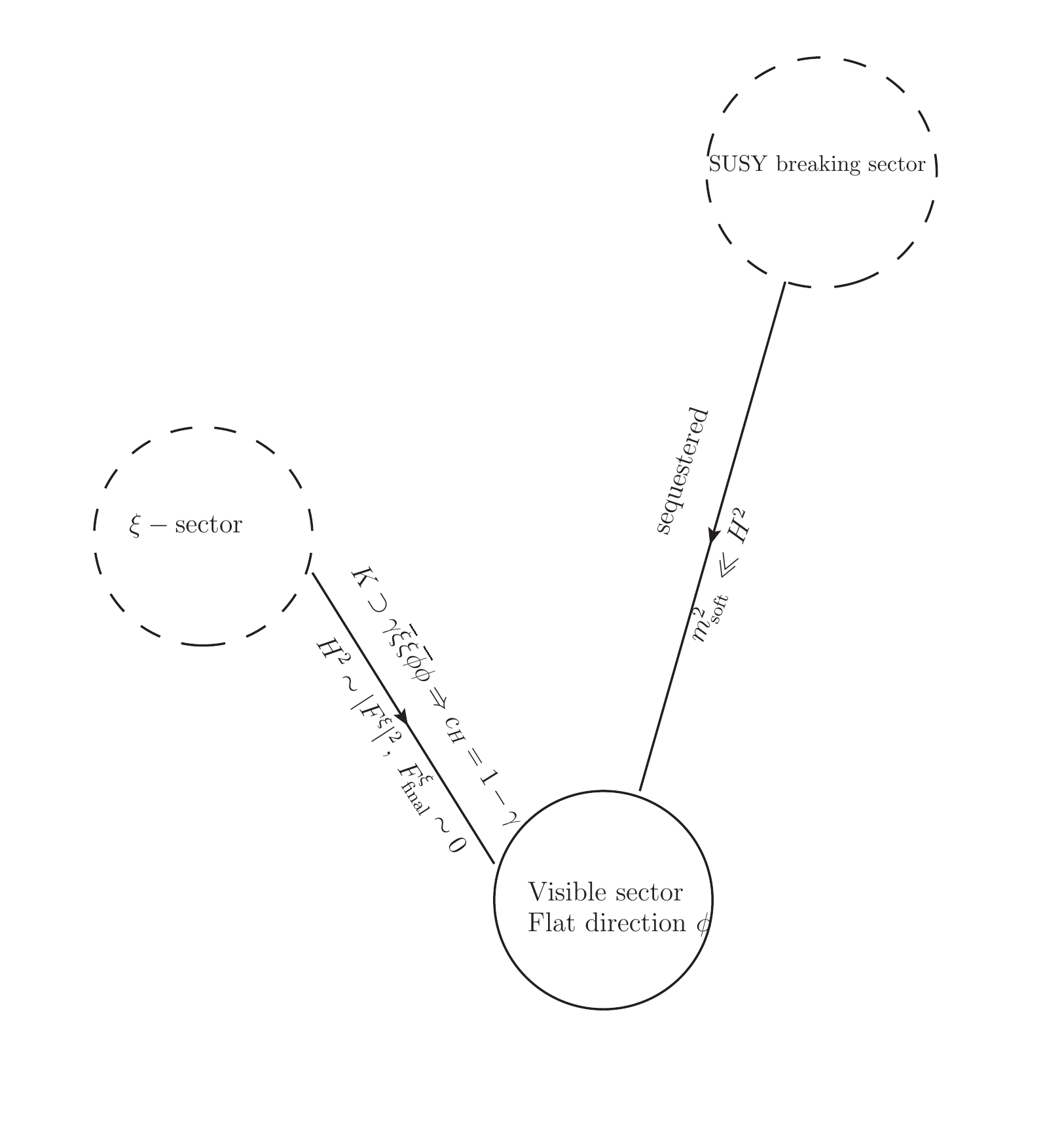}
\caption{The role of various hidden sectors for Affleck-Dine baryogenesis and supersymmetry breaking. The $\xi$-sector dominates the energy density during inflation and induces tachyonic masses along visible sector flat directions. In the final vacuum, the $\xi$-sector contributes negligibly to soft masses, which are sourced by a sequestered supersymmetry breaking sector.}
\end{figure}

We first outline this scenario in type IIB.

\subsection{Type IIB}

It is interesting to study cases when supersymmetry breaking effects during inflation in the hidden matter sector have only mild effects on the moduli sector. We add to the moduli stabilization sector a hidden sector where supersymmetry is broken independently of gravitational effects at an intermediate scale $\mu$ \cite{Kallosh:2006dv, Dine:2004is, Abe:2006xp, Lebedev:2006qq, Dudas:2006gr}.

Concretely, we take
%
%

%
\bea
K \, &=& \, \widehat{K}(T_i) + K_{{\rm hidden}}(\xi) + \widetilde{K}(T_i, \xi)\overline{\phi}\phi  \nonumber \\
\widetilde{K}(T_i, \xi) \, &=& \, \frac{1}{\mathcal{V}^{2/3}} (1 + \gamma \, \overline{\xi} \xi) \nonumber \\
W \, &=& \, \widehat{W}(T_i) + W_{{\rm hidden}}(\xi) \,.
\eea
Here, $\xi$ is a hidden sector matter field.

If the F-term for $\xi$ dominates during inflation, we obtain for $\xi \ll 1$
\be
c_H \sim 1 - \gamma \,\, .
\ee
For $\gamma > 1$, it is possible to obtain a negative induced mass during inflation. However, note that the soft scalar masses in the final vacuum of the theory, in the case of (hidden) matter domination, are given by
\bea
m^2 \, &\sim &\, m_{3/2}^2 - \gamma \left | F^{\xi} \right |^2 \nonumber \\
&\sim &\, m_{3/2}^2 (1 - 3 \gamma) \,\,.
\eea
This leads to tachyons for $\gamma > 1$.

To avoid tachyons and couplings that give rise to the flavor problem, the final supersymmetry breaking should not be matter dominated but sourced by another (sequestered) sector.

A concrete model of inflation depends on the moduli and hidden sector superpotentials. For the modulus, racetrack superpotentials can be appropriately tuned to give inflection or saddle point inflation \cite{BlancoPillado:2006he},
\be \label{modulusfinal}
W(T) \, = \, W_{\rm flux} \,+\, Ae^{-aT} \,+\, Be^{-bT}
\ee
with $A,B \, \sim \, 1$ and $a = 2\pi/N_a, b=2\pi/N_b$ where $N_a,N_b$ are the ranks of the gauge groups on the D7-branes.

%
%
For the hidden sector, one may choose a model that breaks supersymmetry globally. Inflation with Polonyi or O'Raifeartaigh models \cite{Aharony:2007db}, \cite{Franco:2007ii} with $W_{{\rm hidden}} \, = \, \xi_0 \, + \mu^2 \xi$ have been studied recently in this context \cite{Badziak:2009eh}. One has $F^{\xi} \gg F^T$ during inflation, and thus for $\gamma > 1$ one obtains acceptable conditions for AD baryogenesis. However, the final vacuum also has matter dominated supersymmetry breaking, leading to tachyons in the visible sector.

For the purpose of baryogenesis, it is thus more appropriate to choose hidden sector models that have supersymmetry preserving vacua. To lowest order in $\xi/M_P$, the total potential can be written as
\be
V \, = \, e^{\widehat{K}}V(\xi) \, + \, V(T)\,\, .
\ee
An interesting possibility arises if $\xi$ is a pseudo-modulus. Suppose there is a supersymmetry preserving vacuum at $\xi = \xi_{\rm susy}$. For $\xi \, \ll \, \xi_{\rm susy}$, where the supersymmetry restoration effects are subdominant, $\xi$ is flat at tree-level. In this regime, $V(\xi)$ is generated at one-loop and may be suitable for inflation with $F^{\xi} \gg F^T$. Inflation ends in the vacuum at $\xi_{\rm susy}$.

Although the situation outlined above is quite generic, we may consider the concrete example of SQCD in the free magnetic range \cite{Intriligator:2006dd}, \cite{Essig:2007xk}, \cite{Intriligator:1995au}. At low energies, the theory is given by a weakly coupled $SU(N)$ gauge theory with $N_f$ magnetic quarks $q_i$ and a gauge singlet $N_f \times N_f$ meson $\xi$. The tree-level superpotential and Kahler potential are given by
\bea
W(\xi) \, &=& hq_i \xi ^i_j \tilde{q}^j - h \mu^2 \xi_i^i \nonumber \\
K(\xi) \, &=& \xi^{\dagger}\xi + \tilde{q}^{\dagger}\tilde{q} + q^{\dagger}q \,\,.
\eea
Note that $\mu \ll M_P$ and $h$ is dimensionless. The pseudo-modulus piece is a $N_f-N$ block of $\xi$; however, for simplicity, we will refer to this direction as $\xi$. Supersymmetric vacua are obtained by competition between the tree-level piece and a non-perturbative contribution to the superpotential generated by $SU(N)$ gaugino condensation
\be
W_{np} \, = \, N(h^{N_f} \Lambda_m^{-(N_f-3N)} {\rm det} \xi)^{1/N} \,\,,
\ee
where $\Lambda_m$ sets the scale of the IR free theory above which it is strongly coupled. One has $\mu  \ll \xi_{\rm susy} \ll \Lambda_m \ll M_P$. For $\xi \ll \mu$, Coleman-Weinberg corrections give a quadratic potential and a metastable supersymmetry breaking vacuum at the origin. In the intermediate region $\xi \sim \mu$, the corrections are logarithmic. Thus, one has in this regime
\be
V \, \sim \, \frac{1}{(T+\overline{T})^3} \mu^4 \ln (|\xi |^2) \, + V(T) \,\,.
\ee
In this regime, the potential along $\xi$ is extremely flat, sloping very gently toward $\xi_{\rm susy}$, and it is possible to implement inflation as suggested in \cite{Intriligator:2006dd} (note that the modulus sector is also at a saddle/inflection point by choice of parameters). To obtain acceptable inflation, a suitable choice of parameters $(W_{\rm flux}, A, B, a, b)$ in the modulus superpotential and $\mu$ in the matter potential is required, which we leave for future work. The scale of inflation is given by $H \sim \mu^2$, with $F^\xi$ dominating. For $\gamma > 1$, the field $\xi$ induces tachyonic masses along visible sector flat directions. Inflation ends with $\xi$ rolling out to $\xi_{\rm susy}$, restoring supersymmetry. If the racetrack superpotential is also chosen such that the final vacuum along $T$ is Minkowski, this effectively decouples the scale of inflation from constraints of moduli stabilization. One can thus choose $H \sim \mu^2 \gg m_{3/2}$.

We thus have a scenario where
\bea
c_H &=& 1- \gamma \sim -1, \,\, H \gg m_{\rm soft} \, , \nonumber \\
m_{\rm soft} \, &\sim & 100 {\rm GeV},
\eea
where the soft mass along the flat directions is induced by a sequestered sector, while the soft mass induced by the $\xi-$sector vanishes since $F^{\xi} \sim 0$. The flat directions are thus able to acquire non-zero vev during inflation and AD baryogenesis can proceed.

\subsection{Matter-domination in Heterotic Models}

We can  obtain tachyonic Hubble-induced masses in weakly coupled heterotic models. If the energy density during inflation is dominated by the F-term of a matter field $\psi$, then a coupling like $\gamma \overline{\psi}\psi \overline{\phi} \phi$ can induce negative mass along the flat direction $\phi$ similar to the type IIB case. In this case, one has to ensure that the dilaton and Kahler moduli are stabilized and dominate the final supersymmetry breaking in the stable vacuum of the theory.

In modular invariant theories, the scalar potential has stable minima at $T_i = 1, {\rm exp}(i\pi/6)$. At this vacuum, assuming that matter fields vanish, the dilaton can also be stabilized by appropriately choosing the parameters of the non-perturbative contribution $g(l)$ in Eq.~(\ref{Khet1}). To obtain conditions for inflation, note that the superpotential in Eq.~(\ref{Whet}) can be taken, for example, to be
\be
W \, = \, \lambda \psi_1 \eta(T_2)^{-2} \eta(T_3)^{-3} \,\,.
\ee
Then, assuming the vev of $\psi_1 = 0, \,W_{\psi_1} \neq 0$ and that the F-terms of all other fields vanish, the scalar potential can be shown to be independent of untwisted matter in the first moduli sector $T_1$ at tree level \cite{Kain:2006nx}. Such a matter field belonging to the first sector can be taken as the inflaton.
\be
	V_{\rm tree}=\frac{\ell e^g}{1+b\ell}\frac{|\lambda|^2}{x_2 x_3|\eta_2\eta_3|^4 }
\ee
The inflaton dependence enters at loop level. Thus, we have a situation where moduli are fixed and inflation may be obtained with the vacuum energy being dominated by a matter field $\psi_1$. The final vacuum of the theory is dominated by F-terms of moduli and the dilaton. Appropriate coupling of $\psi_1$ to the visible sector flat directions leads to an implementation of AD baryogenesis.

\section{Conclusion}

In this paper, we have investigated Affleck-Dine baryogenesis in a variety of supergravity scenarios. Since the Affleck-Dine process naturally produces large baryon asymmetry, it is useful in obtaining the correct BAU even if there are late-decaying moduli which dilute previously existing asymmetry.

However, successful AD baryogenesis requires tachyonic masses along supersymmetric flat directions during inflation. If we consider such masses to be provided by the F-term of a modulus $\sigma$, then to avoid tachyons in the final vacuum the final F-term of the modulus should vanish.

Whether or not $\sigma$ is able to produce tachyonic masses along flat directions depends on an interplay between global and local geometric data. The details of inflation, the energy density, depend on global data such as the Kahler potential of the moduli fields. The induced soft masses also depend on local data, such as the Kahler metric of the chiral superfields in the visible sector.

We have considered the separate cases of when $\sigma$ is a geometric modulus, and when it is a hidden sector scalar in type IIB. For a geometric modulus, we have argued that the Hubble-induced masses are generically positive if the modulus $\sigma$ is a non-local one. If it is a local modulus and the visible sector is constructed on cycles larger than the string scale, then the induced masses are typically positive. If the construction is at a singularity, then the induced mass depends on mirror type IIA variables, and may become negative for appropriate choice of fluxes on the mirror manifold.

On the other hand, if the modulus $\sigma$ is a hidden sector scalar, then it is possible to satisfy conditions for AD baryogenesis through higher-order couplings of $\sigma$ to the visible sector. We have outlined a scenario which satisfies all the requirements, based on a hidden sector that breaks supersymmetry at an intermediate scale during inflation, but has supersymmetry preserving final vacua.

We have also studied the two cases in global constructions of the weakly coupled heterotic string. The results are similar to type IIB. If $\sigma$ is a geometric modulus, then it may be possible to obtain tachyonic Hubble-induced masses if the flat directions lie along twisted sectors with oscillators. If it is a matter scalar, then it is possible to construct inflationary models where the energy density is dominated by the matter, and appropriate coupling to the visible sector induces tachyonic Hubble-induced mass.

\section{Acknowledgement}

We thank Brent Nelson and especially Rouzbeh Allahverdi for helpful correspondence and discussions. The work of B.D. is supported in part by the DOE grant DE-FG02-95ER40917 and the work of K.S. is supported by NSF under grant PHY-0505757, PHY05-51164.


\end{document}